\documentclass[twocolumn,showpacs,superscriptaddress,amssymb,10pt,pra,floatfix]{revtex4-1}

\usepackage{graphicx}% Include figure files
\usepackage{dcolumn}% Align table columns on decimal point
\usepackage{bm}% bold math
\usepackage{epsfig}
\usepackage{xcolor}
\usepackage{longtable}
\usepackage{hyperref}
\hypersetup{hidelinks}
\usepackage{enumerate}
\usepackage{ulem}
\usepackage{amsmath}

\newcommand{\p}{\partial} 				% partial derivatives
\newcommand{\e}{\mathrm{e}}				% Euler's number
\renewcommand{\i}{i}					% imaginary unit
\renewcommand{\d}{\mathrm{d}}			% differentials
\renewcommand{\vec}{\bm}				% vectors as bold characters 
\renewcommand{\emph}[1]{\textit{#1}}

\newcommand{\order}{\mathcal{O}}

\newcommand{\transm}{\xi}

\newcommand{\reworked}[1]{#1}
\newcommand{\approved}[1]{#1}

\begin{document}

\preprint{}
%
%
%-----------------------------------------Title---------------------------------------
%
%
\title{%\approved{Using Earth's Gravity for Laser Frequency Stabilization}
	\reworked{Using gravitational light deflection in %high-finesse 
		optical cavities for laser frequency stabilization}}
%
%
%-----------------------------------------Author--------------------------------------
%
%

\author{S.~Ulbricht}
\email{sebastian.ulbricht@ptb.de}
\affiliation{Physikalisch--Technische Bundesanstalt, D--38116 Braunschweig, Germany}
\affiliation{Technische Universit\"at Braunschweig, D--38106 Braunschweig, Germany}

\author{J.~Dickmann}
\email{johannes.dickmann@ptb.de}
\affiliation{Physikalisch--Technische Bundesanstalt, D--38116 Braunschweig, Germany}
\affiliation{Technische Universit\"at Braunschweig, D--38106 Braunschweig, Germany}
\affiliation{LENA Laboratory for Emerging Nanometrology, D--38106 Braunschweig, Germany}

%\author{S.~Kroker}
%\affiliation{Physikalisch--Technische Bundesanstalt, D--38116 Braunschweig, Germany}
%\affiliation{Technische Universit\"at Braunschweig, D--38106 Braunschweig, Germany}
%\affiliation{LENA Laboratory for Emerging Nanometrology, D--38106 Braunschweig, Germany}

\author{A.~Surzhykov}
\affiliation{Physikalisch--Technische Bundesanstalt, D--38116 Braunschweig, Germany}
\affiliation{Technische Universit\"at Braunschweig, D--38106 Braunschweig, Germany}
\affiliation{LENA Laboratory for Emerging Nanometrology, D--38106 Braunschweig, Germany}

\date{\today}

%
%
%-----------------------------------------Abstract--------------------------------------
%
%

\begin{abstract} 
We theoretically investigate the propagation of light in the presence of a homogeneous gravitational field.
	To model this, we derive the solutions of the wave equation in Rindler spacetime, which account for gravitational redshift and light deflection.
	The developed theoretical framework is used to explore the propagation of plane light waves in a horizontal Fabry-Perot cavity.
	We pay particular attention to the cavity output power. 
	It is shown that this power depends not only on the input frequency, but also on the vertical position of a detector. 
	We state that the height-dependent detector signal arising from the cavity internal light deflection effect (CILD-effect) also opens a new alternative way to frequency stabilization in Earth-based laser experiments and to study gravitational light deflection at laboratory scales.
\end{abstract}
\maketitle
%
%
%\newpage
\section{Introduction} \label{sec:content_light} \label{page:laser_light_intro}
% history of light deflection
The effect of gravitational light deflection is commonly observed at large scales.
For instance the light of distant stars is known to be bent by the sun's gravitational potential. 
As viewed from Earth, this results in a shift of the apparent position of stars close to the sun, which naturally can only be observed when the solar disc is obscured. The first observation of this phenomenon was made by Eddington during a solar eclipse in 1919, confirming one of the groundbreaking predictions of Einstein's theory of general relativity \cite{Einstein1916,Eddington1919,Book:Wald1984}. Since then, further observations of gravitational lensing \cite{Walsh1979}, microlensing \cite{Refsdal1964,Irvin1989}, and the direct imaging of black holes \cite{Akiyama2019,Akiyama2022} confirmed the existence of gravitational light deflection at galactic up to cosmological scales.  

% light deflection observation in Earth-based experiments, especially in cavities
Despite these astrophysical observations, there is still no small-scale verification of the gravitational light deflection effect. However, advances in optical instruments such as gravitational wave detectors \cite{Harry2010,Sathyaprakash2012,Dooley2015,Baker2019,ET2020}
and highly stable optical resonators \cite{Kessler2012,Matei2017} 
can provide a way to study light deflection in laboratory experiments.  Recently we have proposed to investigate this effect by employing high-finesse optical cavities, as used for frequency stabilization in state-of-the-art optical experiments \cite{Ulbricht2020-1,Ulbricht2020-2,Ulbricht_Thesis2022}. 
%\textcolor{red}{[acknowledge related literature concerning cavities in gravity or freely propagating light on Earth \cite{Richard2019,Raetzel2018,DiPumpo2022}]}.
In such a cavity, light oscillates between two highly reflective mirrors and remains confined for several milliseconds.
For this time the light is exposed to the Earth's gravitational field and literally falls down into vertical direction.
The signatures of this cavity internal light deflection effect (CILD-effect) can be seen in the cavity output signal.

The experimental verification of the CILD-effect would be of great value for research in fundamental physics, since it would 
allow to study the interaction of Earth’s gravity and light in a reproducible way and under well-controlled laboratory conditions. In this work, we present a measurement scheme, that allows not only to verify the effect but also to consider it's possible applications. 
This scheme relies on the application of a vertically segmented detector that analyzes the cavity output signal at two different heights. For such a detector the CILD-effect generates a signal that is highly sensitive to frequency variations near the cavity resonance. This implies that such a signal could be used as the basis for a new laser frequency stabilization method that makes advantage of the Earth’s gravitational potential.
%
% end.

% new paragraph
To investigate light deflection in an optical cavity, we fist have to describe the propagation of light in the Earth's gravitational field. In Sec.~\ref{sec:Rindler_spacetime} we show how this field at laboratory scales can be approximated by Rindler spacetime, i.e., the spacetime of homogeneous acceleration.
The covariant Maxwell equations in this spacetime are used in Sec.~\ref{sec:light_propagation} to obtain a wave equation, which accounts for gravitational redshift and light deflection as the leading order gravitational effects.	
The obtained equations are used in Sec.~\ref{sec:content_cavities_1} to study the propagation and reflection of light between the two highly reflective mirrors of a Fabry-Pérot cavity. 
In particular, we analyze how the cavity output signal depends on the input laser frequency and the vertical position of a detector. Based on the result of our analysis we propose in  Sec.~\ref{sec:frequency_stabilization}  to use a segmented detector which compares the cavity output signal in it's upper and lower segment.
We argue that the power difference at the detector segments is highly sensitive to frequency variations near the cavity resonance frequency and, thus, can open new ways towards the development of novel frequency stabilization techniques, as well as to explore gravitational light deflection at laboratory scales.
%
% end.

\section{Maxwell equations in Rindler spacetime} \label{sec:Rindler_spacetime}
	In the present work, we will investigate the effect of gravity on the propagation of light in optical cavities.
	Theoretical basis of this study is given in terms of Maxwell equations, formulated in a gravitationally distorted spacetime.
	Since we discussed this theoretical framework in detail in our former publications \cite{Ulbricht2020-1,Ulbricht2020-2}, here we restrict ourselves to a brief compilation of the basic formulas and ideas, formulated in the language of general relativity.
	% end.

% new paragraph
	For typical laboratory experiments on Earth, spacetime can be described by the Rindler line element  
	\begin{equation}
		\d s^2 = g_{\mu\nu}\d x^\mu \d x^\nu =\left(1+\frac{gz}{c^2}\right)^2\,\d(ct)^2-\d \vec{r}^2\, \label{eqn:Metric}
	\end{equation}
	to good approximation. 
	This line element accounts for the approximately homogeneous gravitational acceleration  $\vec{g}=-g\,\vec{e}_z$ with $g= 9.81\, \mathrm{m}/\mathrm{s}^2$ on our planet's surface \cite{Rindler1960, Rindler1966}.
	Additional effects according to the Earth as a spherical body can be neglected in a small region around an observer's position.
	In the expression above $c$ is the speed of light and $g_{\mu\nu}$ is the metric tensor with the sign-convention $(1,-1,-1,-1)$ in Cartesian coordinates with $\vec{r}=(ct,x,y,z)$.  
	As usual in relativistic physics, we use the Einstein notation in which a summation  from 0 to 3 is performed when paired Greek letters appear.
	%
	% end.

% new paragraph
	The Maxwell equations in Rindler spacetime can be written in the covariant form
	\begin{equation}
		\nabla_\mu F^{\mu\nu}=0\,, \label{eqn:maxwell}
	\end{equation} 
	where $F^{\mu\nu}$ is the electromagnetic field strength tensor, whose components are related to the usual electric and magnetic fields \cite{Book:Wald1984,Book:Carrol2004}.
	For our further theoretical analysis, it is convenient to construct this tensor $F_{\mu\nu}=\p_\mu A_\nu-\p_\nu A_\mu$ from the partial derivatives of the four-potential $A_{\mu}=(\,\Phi/c\,,\-\vec{A}\,)$. Here $\vec{A}$ and $\Phi$ are the electromagnetic vector and scalar potentials of the electromagnetic fields.
	Moreover, in Eq.~(\ref{eqn:maxwell}) the covariant derivative $\nabla_\mu=\p_\mu+\Gamma^{\rho}_{\mu\rho}$ is constructed from the partial one $\p_\mu$ and from the Christoffel symbol $\Gamma^\rho_{\mu\rho}=g/c^2 \, \left(1+gz/c^2\right)^{-1}\delta^{3}_\mu$, which contains the geometric information about Rindler spacetime.
	%
	% end.

% new paragraph
	By inserting the four-potential $A_{\mu}$ into the field strength tensor $F^{\mu\nu}$ and employing Eq.~(\ref{eqn:maxwell}) we can derive the covariant wave equations
	\begin{eqnarray}
		\nabla_\mu\nabla^\mu \vec{A}=0\,,\quad \nabla_\mu\nabla^\mu \Phi=0\,,\label{eqn:covwave}
	\end{eqnarray}
	where the Lorentz gauge condition  $\nabla_{\mu}A^{\mu}\!\!=\!0$ is assumed.
	For propagating electromagnetic fields this condition implies that the scalar potential $\Phi$ can be derived from the vector potential using the relation
	\begin{equation}
		\p_t\Phi/c=\left(1+\frac{g z}{c^2}\right) \vec{\nabla} \!\cdot\!\left[\left(1+\frac{g z}{c^2}\right) \vec{A}\right]\,. \label{eqn:Lorentz_gauge}
	\end{equation}
	In the next section, therefore, we will restrict our analysis to the vector potential $\vec{A}$, which will be utilized to describe the propagation of light in the homogeneous gravitational field.
	%
	% end.

\section{Light propagation in a homogeneous gravitational field} \label{sec:light_propagation}
	In order to proceed with our theoretical analysis, we need to simplify the wave equation for the vector potential $\vec{A}$, given in Eq.~(\ref{eqn:covwave}). In particular we want to account for the leading order effects of gravity on light propagation on Earth. For that purpose, in Eq.~(\ref{eqn:covwave}) we use the explicit form of the covariant derivatives and perform an expansion in the small parameter $\epsilon=gL/c^2$, where $L$ is a typical length scale of the experiment.
	The resulting wave equation then has the form
	\begin{equation}
		\frac{1}{c^2}\frac{\p^2}{\p t^2}\vec{A} -\left[\left(1+\frac{g z}{c^2}\right)\vec{\nabla}\right]^2\vec{A}=0 + \order{}\left(\epsilon^2\right)\,,\label{eqn:wave_equation}
	\end{equation}
	which looks quite similar to the conventional wave equation, but differs from that by the prefactor $(1+gz/c^2)$ in front of the usual nabla operator $\vec{\nabla}=(\p_x,\p_y,\p_z)$. This prefactor will give rise to various effects, such as gravitational redshift and light deflection.
	In what follows, we will use the solutions to Eq.~(\ref{eqn:wave_equation}) to discuss the impact of Earth's gravity on freely propagating light in the homogeneous gravitational field.
% end.

\subsection{A set of basis functions for the electromagnetic vector potential} \label{sec:basis_functions}
	As in the case of the conventional wave equation, the solution of Eq.~(\ref{eqn:wave_equation}) can be found in terms of a superposition of basis functions.
	In what follows, we will choose a basis set of paraxial solutions for which the momentum of light is concentrated along the $x$-axis, i.e., $k_y,k_z\ll k_x$.
	In this regime, the amplitude and polarization degrees of freedom decouple and the full vector potential can be expressed by the superposition
	\begin{equation}
		\vec{A}(\vec{r},t)=\int\!\!\!\!\int\!\!\!\!\int\!\vec{e}\,\tilde {A}_{{\vec{k}}}\psi^{\vec{k}}_{\delta k_z}(\vec{r},t)\, \d (k_z^2)\d k_y \d k_x\,. \label{eqn:superposition}
	\end{equation}
	Here, $\vec{e}$ is a constant polarization vector and $\psi^{\vec{k}}_{\delta k_z}(\vec{r},t)$ are the scalar basis functions, solving the wave equation Eq.~(\ref{eqn:wave_equation}) for any chosen parameter vector $\vec{k}=(k_x,k_y,k_z)$, which obeys the dispersion relation $\vec{k}^2=(2\pi\nu)^2/c^2$.
	These basis functions, whose derivation has been discussed in detail in  Ref.~\cite{Ulbricht2020-2}, are given by
	\begin{eqnarray}
		&&\psi^{\vec{k}}_{\delta k_z}(\vec{r},t)\label{eqn:basis_functions}\\&&=\frac{1}{2\pi}\frac{1}{\sqrt{\delta k_z}}\e^{-\frac{gz}{2c^2}}\mathrm{Ai}\left[-\frac{k_z^2}{\delta k_z^2}+\delta k_z z\right]\!\e^{ik_x x}\,\e^{ik_y y}\,\e^{-2\pi i \nu t}\,,\nonumber
	\end{eqnarray} 
	where $\delta k_z = 2(g\pi^2\nu^2/c^4)^{1/3}$ is a small parameter and $\mathrm{Ai}(x)$ is the Airy function \cite{Book:Vallee2004} that arises from the properties of the wave equation Eq.~(\ref{eqn:wave_equation}) along the vertical direction.
	% end.

	%new paragraph
	One can recognize the unconventional parameter space $\d (k_z^2)\d k_y \d k_x$ used in the superposition integral Eq.~(\ref{eqn:superposition}). 
	This is due to the fact that the basis functions  Eq.~(\ref{eqn:basis_functions}) satisfy the completeness relation 
	\begin{eqnarray}
		& &\int\!\!\!\!\int\!\!\!\!\int\psi^{\vec{k}}_{\delta k_z}(\vec{r},t)\psi^{\ast\vec{k}}_{\delta k_z}(\vec{r}',t)\,\d (k_z^2)\d k_y \d k_x\label{eqn:completeness_relation}\\& &\hspace{9em} = \left(1+\frac{gz}{c^2}\right)^{-1}\!\!\delta^{(3)}(\vec{r}-\vec{r}')\,,\nonumber
	\end{eqnarray} 
	cf. Ref.~\cite{Hunt1981}, where the prefactor $(1+gz/c^2)^{-1}$ of the delta function compensates the gravitational modifications to the infinitesimal volume $(1+gz/c^2)\d x \d y \d z$ of Rindler spacetime.
	In the next section, we will use this complete set of basis functions to calculate the vector potential of a gravitationally distorted plane wave.

\subsection{Plane waves in Earth's gravity}
\label{sec:plane_waves}
In order to make use of the basis functions Eq.~(\ref{eqn:basis_functions}) and to describe light propagation in the homogeneous gravitational field, we have to specify the properties of light by a proper boundary condition. In the present study, we will consider a monochromatic plane wave, initially entering a horizontal Fabry-Perot cavity at $x=0$. This implies the boundary condition 
	\begin{eqnarray}
		&&\vec{A}_{\mathrm{in}}(z,t)\label{eqn:boundarycondition}\\&&\quad=\vec{A}_0 \left(1+\frac{gz}{c^2}\right)^{\alpha-1/2}\!\!\! \exp\!\left[-2\pi i  \nu t\left(1+\frac{gz}{c^2}\right)^{-\beta} \right]\,, \nonumber
	\end{eqnarray}
	with a constant amplitude $\vec{A}_0$ and a constant frequency $\nu$.
	For vanishing gravitational acceleration, i.e. $g\to 0$, this boundary condition becomes $\vec{A}_0 \exp(-2\pi\i\nu t)$, resembling the familiar expression for a plane wave at $x=0$.
	In the presence of gravity, however, both the amplitude and the phase of the plane wave are modified by the factor $(1+gz/c^2)$ to certain powers $\alpha-1/2$ and $-\beta$, respectively. While the parameter $\alpha$ is adjustable for technical purposes as mentioned in Appendix \ref{sec:appendix_A}, the parameter $\beta$ accounts for the initial gravitational frequency redshift of the used laser light.
	%
	% end. 

% new paragraph
For the further analysis of light propagation in the homogeneous gravitational field it is practical to expand the boundary condition  Eq.~(\ref{eqn:boundarycondition}) in terms of the basis functions (\ref{eqn:basis_functions}). This expansion reads  
	\begin{equation}
		\vec{A}_{\mathrm{in}}(z,t)=\int\!\!\!\!\int\!\!\!\!\int\!\,\tilde {\vec{A}}^b_{{\vec{k}}}\,\psi^{\vec{k}}_{\delta k_z}(\vec{r},t)\, \d (k_z^2)\d k_y \d k_x\,, \label{eqn:superposition_boundary}
	\end{equation}
	where the expansion coefficients are given by
	\begin{subequations} \label{eqn:boundary_all}
		\begin{eqnarray}
			\tilde {\vec{A}}^{b}_{\vec{k}}&=&\!\!\!\int\!\!\! \vec{A}_{\mathrm{in}}(z',t)\psi^{\ast\vec{k}}_{\delta k_z}(\vec{r}',t)\left(1+\frac{gz'}{c^2}\right) \d x' \d y' \d z'\\
			&=&2\pi \vec{A}_0 \delta(k_x)\delta(k_y)\label{eqn:boundary} \\
			& & \times\frac{1}{\sqrt{\delta k_z}}\int\!\! \d z' \e^{gz'b(\alpha,\beta)/c^2}\mathrm{Ai}\left[-\frac{k_z^2}{\delta k_z^2}+\delta k_z z'\right]\,. \nonumber
		\end{eqnarray}
	\end{subequations}
	\ignorespacesafterend Here, Eq.~(\ref{eqn:boundary}) arises from the explicit form of $\vec{A}_{\mathrm{in}}(z,t)$ and $\psi^{\ast\vec{k}}_{\delta k_z}(\vec{r},t)$ after performing the $x',y'$-integration, while the $z'$-integral is kept for a later treatment.
	Moreover, we introduced the notation  $b(\alpha,\beta)=\alpha+2\pi\i\beta \nu t$ for the boundary condition.
%
% end.

% new paragraph
The expressions (\ref{eqn:superposition_boundary})--(\ref{eqn:boundary_all}) describe the boundary condition at $x=0$. In what follows, however, we want to analyze the propagation of light in an arbitrary point $\vec{r}=(x,y,z)$ with $x>0$. As we discussed in detail in Ref.~\cite{Ulbricht2020-2}, the vector potential $\vec{A}(\vec{r},t)$ at position $\vec{r}$ can be obtained by the replacement 
	\begin{equation}
		\delta(k_x) \to \delta\left(k_x -\frac{2\pi\nu }{c} +\frac{c}{4\pi\nu }(k_y^2+k_z^2)\right)\,, \label{eqn:delta}
	\end{equation}
	in Eqs.~(\ref{eqn:superposition_boundary})--(\ref{eqn:boundary_all}). This implements the dispersion relation $2\pi\nu /c=\sqrt{k_x^2+k_y^2+k_z^2}$
	%\approx 2\pi\nu/c -c(k_y^2+k_z^2)/(4\pi\nu)$ 
	in the paraxial regime, where $k_y,k_z\ll k_x$.
	Making use of relation (\ref{eqn:delta}), the expansion coefficients $ \tilde {\vec{A}}^{b}_{\vec{k}}$ are replaced by their counterparts $\tilde {\vec{A}}_{\vec{k}}$ of the vector potential $\vec{A}(\vec{r},t)$ at an arbitrary position. Thus, we obtain
	\begin{subequations} 
		\begin{eqnarray}
			\vec{A}(\vec{r},t)&=&\int\!\!\!\!\int\!\!\!\!\int\tilde {\vec{A}}_{{\vec{k}}} \psi^{\vec{k}}_{\delta k_z}(\vec{r},t)\, \d (k_z^2)\d k_y \d k_z\\
			&=&\vec{A}_0\,\left(1+\frac{gz}{c^2}\right)^{-1/2}\e^{-2\pi i \nu (t-x/c)}\,\e^{S_g}\label{eqn:rearanged}\\
						&&\hspace{14em}+\order{}(\epsilon^2)\,, \nonumber
		\end{eqnarray}
	\end{subequations}
	where the integration over $k_x$ and $k_y$ can be easily performed employing the delta function (\ref{eqn:delta}). The remaining integrals over $z'$ and $k_z^2$ in Eq.~(\ref{eqn:rearanged}) are collected in the complex exponential 
	\begin{equation}
		\e^{S_g}=\int\d z' \e^{gz'b(\alpha,\beta)/c^2} \, f(x,z,z') \,, \label{eqn:grav_exponent}
	\end{equation}
	and the kernel function
	\begin{subequations}
		\begin{eqnarray}
			& &f(x,z,z')\nonumber\\
			& &=\int \d (k_z^2) \frac{1}{\delta k_z} \e^{-\i \frac{k_z^2 cx}{4\pi\nu }}\mathrm{Ai}\left[-\frac{k_z^2}{\delta k_z^2}+\delta k_z z\right]\label{eqn:Ai_integral}\\
			& &\hspace{12em}\times 
			\mathrm{Ai}\left[-\frac{k_z^2}{\delta k_z^2}+\delta k_z z'\right]\nonumber \\
			&&= \sqrt{\frac{\nu }{\i cx}}\exp\left[\phantom{+}\!\!\!\i \frac{\pi\nu }{cx}(z-z')^2\,-\,\i \frac{gx}{c^2}\frac{\pi\nu }{c}(z+z')\right]\label{eqn:cernal}\\
			&&\hspace{19em}+\order{}(\epsilon^2)\,. \nonumber
		\end{eqnarray}
	\end{subequations}
	Here, the integral (\ref{eqn:Ai_integral}) can be reformulated and solved in terms of
	standard Gaussian integrals
	using the cosine representation of the Airy function \cite{Book:Vallee2004}.
	We note, moreover, that in Eq.~(\ref{eqn:cernal}) only the linear order in $\epsilon=gL/c^2$ is considered.
%
% end.

\begin{figure}[t]
		\centering
		(a)\includegraphics[scale=0.165]{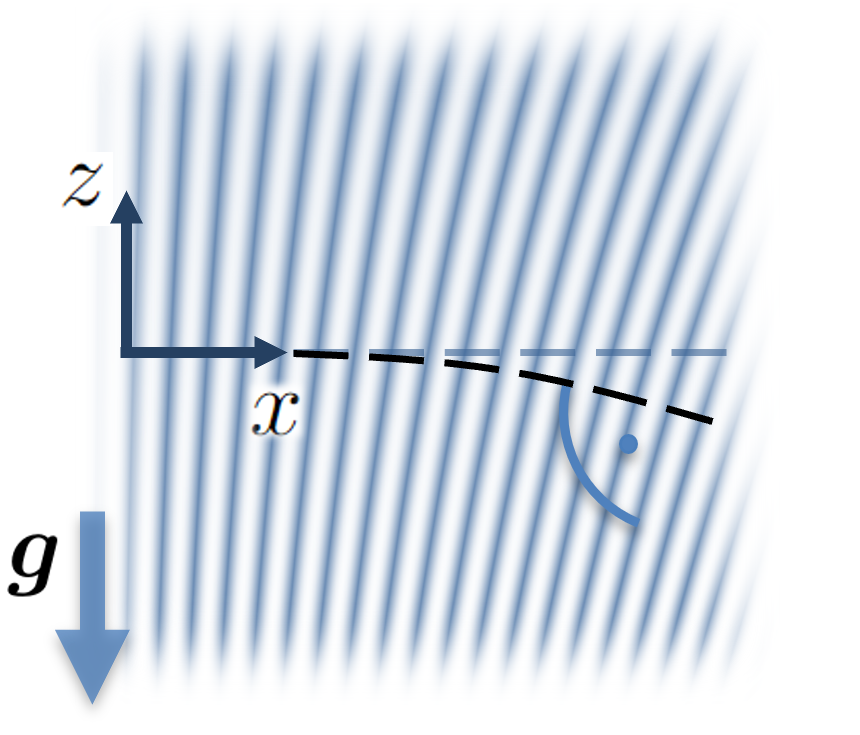}(b)\includegraphics[scale=0.165]{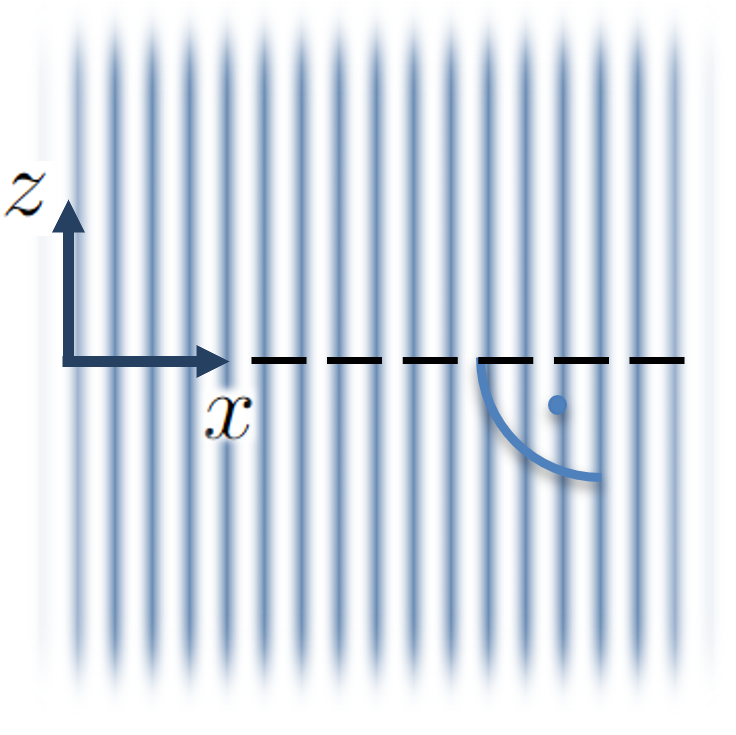}\\ 
		\caption{
			\approved{Illustration of a horizontal propagating monochromatic wave (a) in the homogeneous gravitational field  in comparison to the case (b) of plane waves in the absence of gravity .
				We find that in the presence of gravity the light initially starts as a plane wave at $x=0$ and is bent downwards while its propagation in positive $x$-direction.}} \label{fig:plane_wave}
	\end{figure}

%new paragraph
By inserting the kernel function (\ref{eqn:cernal}) into Eq.~(\ref{eqn:grav_exponent}) we finally obtain the complex exponent
	\begin{equation}
		S_g=\frac{gz}{c^2}b(\alpha,\beta)\,-\,\i \frac{gz}{c^2}\frac{2\pi \nu }{c}\,x\,. \label{eqn:gravity_exponent}
	\end{equation}
	While the first term in this expression contains information about the boundary condition, the second term is proportional to the product $zx$ of the coordinates into the directions of gravity and light propagation. 
	Due to that, the phase fronts of the light wave are deformed and bent downwards with increasing propagation distance.
	This is an expression of the gravitational deflection of the vector potential, as visualized in Fig.~\ref{fig:plane_wave}.
	%
	%
	% end.

	% new paragrpah
	By inserting the complex exponent (\ref{eqn:gravity_exponent}) into Eq.~(\ref{eqn:rearanged}), the vector potential of the plane light wave in a homogeneous gravitational field can be finally written as 
	\begin{equation}
		\vec{A}(\vec{r},t)= \vec{A}_0\left(1+\frac{gz}{c^2}\right)^{-1/2}\!\!\exp\left[\frac{gz}{c^2}\alpha+i\phi(x,z,t)\right]\,
		\label{eqn:rearanged2}
	\end{equation}
	with the phase
	\begin{equation}
		\phi(x,z,t)=\frac{2\pi  \nu }{c}\left(x-ct -\frac{gz}{c^2}x+\frac{gz}{c^2}\beta ct\right)\,.
		\label{eqn:rearanged3}
	\end{equation}
	Comparing this result to Eq.~(\ref{eqn:boundarycondition}), we find that expressions (\ref{eqn:rearanged2})-(\ref{eqn:rearanged3}) can be written
	as the product of the initial vector potential and a phase exponential term
	\begin{equation}
		\vec{A}(\vec{r},t)=\vec{A}_{\mathrm{in}}(z,t)\,\exp\left[\frac{2\pi i  \nu }{c}\left(1+\frac{gz}{c^2}\right)^{-1}\!\!x \right]+\order{}(\epsilon^2)\,, \label{eqn:Art}
	\end{equation}
	where we made use of expansions for $\exp(gz/c^2 )=(1+gz/c^2)+\mathcal{O}(\epsilon^2)$ and $(1-gz/c^2)=(1+gz/c^2)^{-1}+\mathcal{O}(\epsilon^2)$.
	The exponential term in Eq.~(\ref{eqn:Art}) describes the propagation of light in $x$-direction.
	The $z$-dependence in this expression leads to a deformation of the vertical wave fronts and gives rise to a height dependent length scale, namely the redshifted wave length $\lambda_g(z)\!=(1+gz/c^2)c/\nu$ that accounts for the effect of gravitational light deflection.
	We mention, moreover, that the exponential term in Eq.~(\ref{eqn:Art}) depends on the properties of Rindler spacetime and is independent of the initial frequency redshift in Eq.~(\ref{eqn:boundarycondition}). 
	% end.

\section{Output Signal of a Fabry-Pérot Cavity in Gravity} \label{sec:content_cavities_1}
\subsection{Vector Potential at the Cavity Output}
\begin{figure}[t]
		\centering
		\includegraphics[scale=0.1]{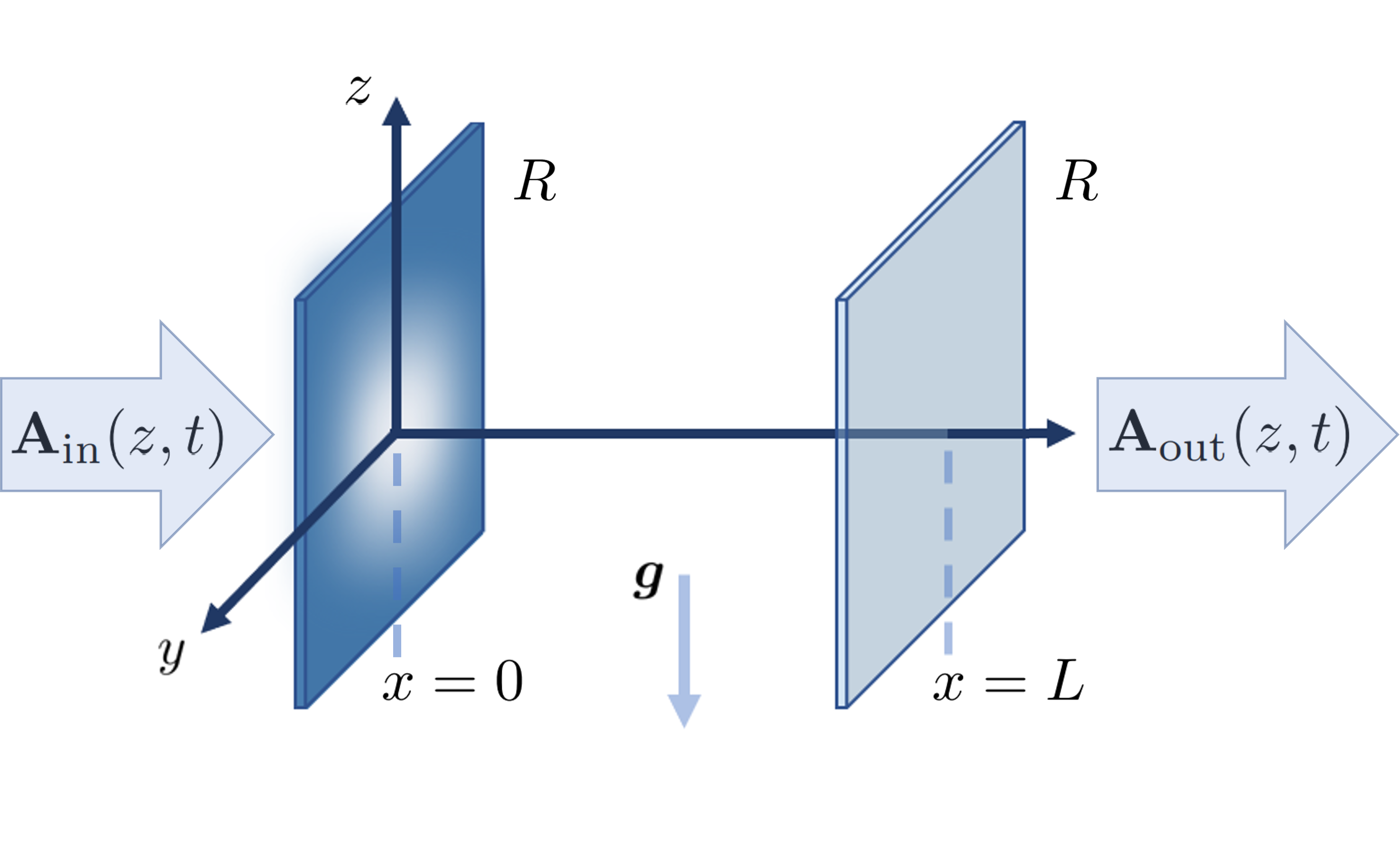}\\ 
		\caption{\approved{Geometry of the horizontal Fabry-P\'erot cavity consisting of two plane-parallel mirrors of reflectivity $R=(1-\transm^2)$. 
				The laser initially enters the cavity at the plane mirror at $x=0$. The cavity output signal is investigated behind a second plane mirror located at $x=L$.}} \label{fig:coplanar_cavity}
	\end{figure}

In the previous section we reviewed a theory for the propagation of a plane monochromatic light wave in the presence of a homogeneous gravitational field.
	Now, we are ready to employ this theory to investigate how gravity affects the measurable outcome of optical cavity experiments.
	In the present work, we consider the particular case of a horizontal Fabry-Pérot cavity consisting of two plane-parallel mirrors at $x=0$ and $x=L$, as shown in Fig.~\ref{fig:coplanar_cavity}.
	The cavity mirrors are assumed to transmit only a small fraction $\transm\ll 1$ of light,
	while the mayor part of light is reflected with an amplitude $(1-\transm) \vec{A}_0$.
	In an idealized scenario, where scattering and absorption of light can be neglected, 
	the amplitude transmittance parameter $\transm$ is directly connected to the common reflectivity  $R=(1-\transm)^2$ \cite{Book:Hecht2002}.
	%
	% end.

%new paragraph
Having briefly discussed the cavity model, we will now describe how light evolves in this optical device.
	For this purpose we use the round trip calculation method, which was discussed for general mirror surfaces in our previous publication \cite{Ulbricht2020-2}. 
	The plane wave with the vector potential (\ref{eqn:Art}) enters the cavity at $x=0$ and propagates towards the second mirror at $x=L$. 
	When the light reaches this second mirror, a small fraction of light is transmitted and constitutes the first contribution to the cavity output signal
	\begin{equation} 
		\vec{A}^{(0)}(z,t)=\transm \,\vec{A}_{\mathrm{in}}(z,t) \,\exp\left[\frac{2\pi i  \nu}{c}\left(1+\frac{gz}{c^2}\right)^{-1}\!\!L \right]\,.
	\end{equation}
	For the other fraction of light that remains in the cavity the amplitude is reduced by a factor $(1-\transm)$ and the propagation direction is changed from $x$ to $-x$ by reflection. Thus, the light propagates back to the first mirror at $x=0$, where it is once again reflected and returns to the output mirror at $x=L$. Back there, the next fraction of light 
	\begin{equation}
		\vec{A}^{(1)}(z,t)=\vec{A}^{(0)}(z,t) (1-\transm)^2 \,\exp\left[\frac{2\pi i  \nu}{c}\left(1+\frac{gz}{c^2}\right)^{-1}\!\!\,2L \right]
	\end{equation}
	leaves the cavity and adds to the cavity output signal. 
	This expression differs from output of the former round trip only by a prefactor $(1-\transm)^2$, which accounts for the two reflections, and a phase that depends on the increase of the propagation distance by $2L$ and the corresponding gravitational redshift.
	By continuation of this procedure we find that the contributions of the $n$-th round trip to the cavity output signal reads 
	\begin{eqnarray}
		\vec{A}^{(n)}(z,t)&=&\vec{A}^{(0)}(z,t) \\
		&&\times\left\{(1-\transm)^2 \,\exp\left[\frac{2\pi i  \nu}{c}\left(1+\frac{gz}{c^2}\right)^{-1}\!\!\,2L \right]\right\}^{n}\!. \nonumber
	\end{eqnarray}
	The contributions of all round trips can be finally summed up to obtain the vector potential at the cavity output
	%\begin{equation}
	%	\vec{A}_{\mathrm{out}}(z,t)=\!\sum\limits_{n=0}^{\infty}\!\vec{A}^{(n)}(z,t)=\frac{\transm \vec{A}_{\mathrm{in}}(z,t) \exp\!\left[\frac{2\pi i  \nu}{c}\!\left(1+\frac{gz}{c^2}\right)^{-1}\!\!\,2L \right]}{1-(1-\transm)^2 \exp\!\left[\frac{2\pi i  \nu}{c}\left(1+\frac{gz}{c^2}\right)^{-1}\!\!\,2L \right]}\,.\!\!\!
	%\end{equation}
	\begin{subequations}
		\begin{eqnarray}
			& &\hspace{-1em}\vec{A}_{\mathrm{out}}(z,t)=\!\sum\limits_{n=0}^{\infty}\!\vec{A}^{(n)}(z,t)\\
			& &\qquad=\frac{\transm\exp\!\left[\frac{2\pi i  \nu}{c}\!\left(1+\frac{gz}{c^2}\right)^{-1}\!\!\,2L \right]\vec{A}_{\mathrm{in}}(z,t)}{1-(1-\transm)^2 \exp\!\left[\frac{2\pi i  \nu}{c}\left(1+\frac{gz}{c^2}\right)^{-1}\!\!\,2L \right]}\,,\quad \label{eqn:AoutzuAin}
		\end{eqnarray}
	\end{subequations}
	where we utilized the geometrical series $\sum_{n=0}^\infty q^n=1/(1-q)$ for $|q|<1$ to perform the summation.
	We find, that the vector potential at the cavity output is related
	to the vector potential of the incident light by a proportionality factor that depends on both parameters of the laser-cavity system ($\nu$, $L$, and $\transm$) and the properties of spacetime ($g$ and $z$).
	This suggests that the cavity output signal carries information about the effect of light deflection and can be used to investigate gravity in an optical experiment on Earth, as we will discuss in the next section.
\subsection{Output intensity and power of a Fabry-Pérot cavity on Earth}\label{sec:output_power}
	In the last section, we obtained the vector potential at the output of a horizontal Fabry-Pérot cavity, as it is displayed in Fig.~\ref{fig:coplanar_cavity}.  In a typical cavity experiment not the vector potential $\vec{A}_{\mathrm{out}}(z,t)$ but the output intensity, which is proportional to $|\vec{A}_{\mathrm{out}}(z,t)|^2$, is measured. Utilizing Eq.~(\ref{eqn:AoutzuAin}) we find this intensity as
	\begin{equation}
		I_{\mathrm{out}}(\nu,z)=\frac{FI_0/4}{1+F \sin^2\left[\frac{2\pi\nu}{c}\left(1+\frac{gz}{c^2}\right)^{-1}L\right]}\,,\label{eqn:output_intensity}
	\end{equation}
	where $I_0=\varepsilon_0 c\,(2\pi\nu)^2 |\vec{A}_0|^2/2$ is the constant intensity of the plane wave, initially entering the cavity.
	We further introduced the parameter $F=4R/(1-R)^2$, which is related to the \emph{cavity finesse} $\mathcal{F}=\pi \sqrt{F}/2$ commonly used to characterize an optical cavity \cite{Ismail2016,Dawkins2007}. For a careful discussion of the definition of intensity in Rindler spacetime, see Appendix \ref{sec:appendix_A}.
%
% end.

% new paragraph
	We can see from Eq.~(\ref{eqn:output_intensity}) that the output intensity is sensitive not only to the variation of the cavity length and the laser frequency, but also depends on the vertical position $z$ at which the output signal is detected, see Ref.~\cite{remark1} for further discussion.

	% new paragraph
	Integrating the intensity Eq.~(\ref{eqn:output_intensity}) over the surface of a rectangular detector with height $2a$ and width $2b$, who's center is located in a specific height $h$, we obtain the collected output power
	\begin{subequations}
		\begin{eqnarray}
			& &P_{\mathrm{out}}(\nu,h)= \int_{h-a}^{h+a}\int_{-b}^{b} I_{\mathrm{out}}(\nu,z) \,\d y \d z\\
			& &\qquad =\frac{AFI_0/4}{1+F \sin^2\left[\frac{2\pi\nu}{c}\left(1+\frac{gh}{c^2}\right)^{-1}L\right]
			}+\mathcal{O}(\epsilon^2)\,,\qquad\label{eqn:output_power}
		\end{eqnarray}
	\end{subequations}
	where $A=4ab$ is the detector surface. Together with Eq.~(\ref{eqn:output_intensity}) this expression implies that the power collected by the detector is proportional to the intensity measured by a point-like detector at a position $h$ in the center of the detector surface:
	\begin{equation}
		P_{\mathrm{out}}(\nu,h)=A\,I_{\mathrm{out}}(\nu,h)\,.
	\end{equation}
	In the next section, we will use these findings and  the expression (\ref{eqn:output_power}) for the output power to discuss a detector scheme that is suitable to measure the impact of gravity on the laser-cavity setup and enables new possibilities for laser frequency stabilization.
%
% end.

\section{Using Earth's gravity for laser frequency stabilization}
\label{sec:frequency_stabilization}
\begin{figure}[b]
		\centering
		\includegraphics[scale=0.18]{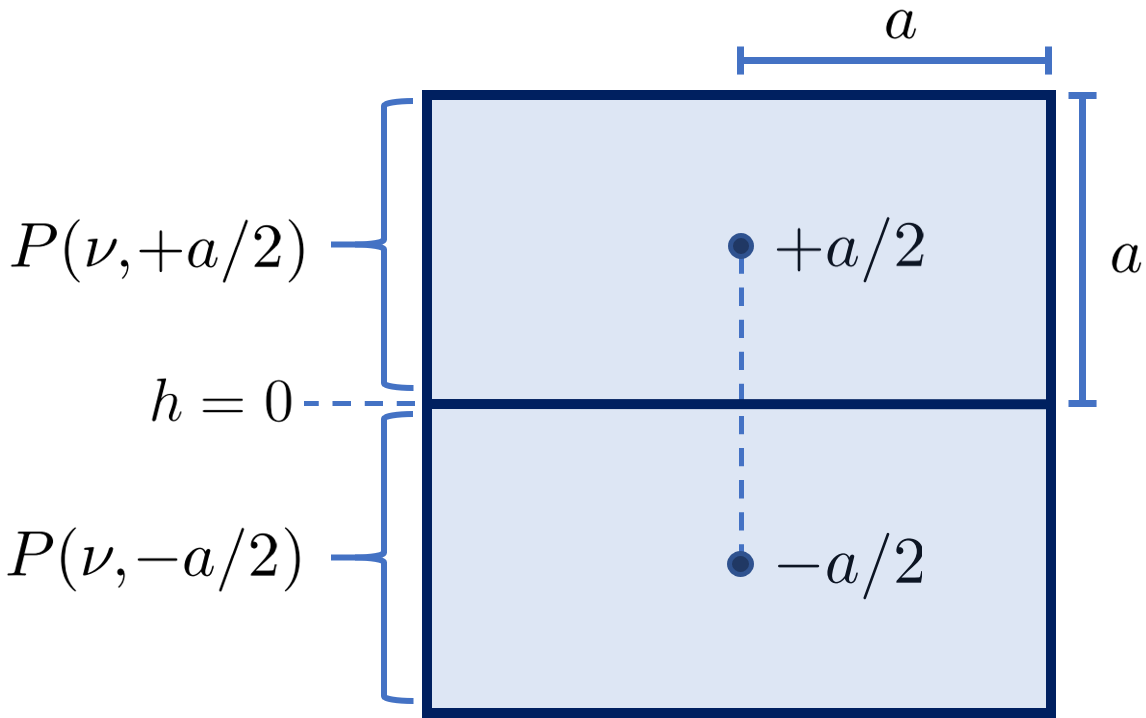}\\ 
		\caption{\approved{Geometry of a segmented detector with two rectangular segments, each of area $2a^2$, on top of each other. The detector behaves like two point-like detectors located at heights $\pm a/2$, collecting the light from the upper and the lower segment, respectively.}} \label{fig:quad_detector_geometry}
	\end{figure}

\begin{figure}[t]
	\centering
	\hspace{-1em}\includegraphics[scale=0.215]{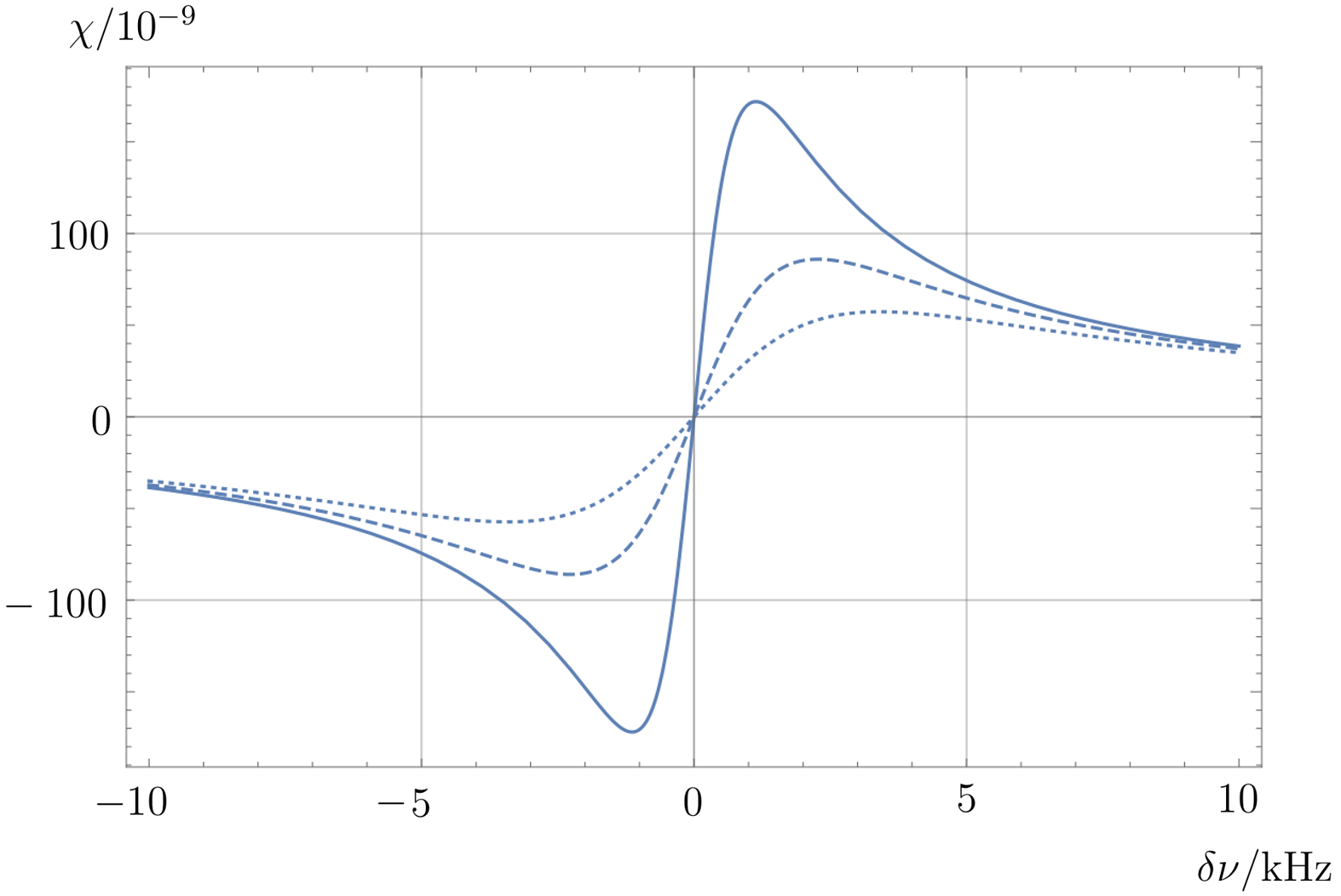}\\ 
	\caption{\reworked{The $\chi$-parameter (\ref{eqn:detector_signal}) as a function of frequency offset $\delta\nu$ with respect to the resonance frequency $\nu_0=2.8\times 10^{14}\,\mathrm{Hz}$ of a $L=21\,\mathrm{cm}$ cavity. The signal for a detector geometry with $a=1.5\,\mathrm{cm}$ (c.f., Fig.~\ref{fig:quad_detector_geometry}) is given for mirror reflectivities of $R=1-3\times 10^{-5}$ (dotted), $R=1-2\times 10^{-5}$ (dashed) and $R=1-10^{-5}$ (solid), respectively.
			In the latter case, regarding to Eq.~(\ref{eqn:photon_noise}), the signal-to-noise ratio can be estimated $\chi/\Delta\chi\sim 200$. }} \label{fig:quad_detector_signal}  
\end{figure}

Thanks to state-of-the-art highly reflective mirrors, optical cavities with finesses of $\mathcal{F}\sim 3\times 10^5$ and beyond can be employed in modern high precision experiments \cite{Kessler2012, Matei2017,Ludlow2007}.
	One of the major applications for such optical devices is the stabilization of laser frequency.
	To shortly recall the concept of laser stabilization we consider Eq.~(\ref{eqn:output_power}) in the case of zero gravity, i.e., $g=0$. As seen from this equation the output power of a high-finesse cavity decreases rapidly, when the light frequency $\nu$ is off the cavity resonance at $\nu_0=n c/2L$, where $n$ is the integer mode number.
	It implies that even only small variations of the frequency lead to big changes of the cavity output power, which can be experimentally observed.
	This effect is a corner stone of the \emph{side-of-fringe} locking technique as one of the established ways to study and compensate laser frequency fluctuations in order to stabilize lasers for high precision applications \cite{Barger1973,Helmcke1982, Huang2014}.
	Another common method, the \emph{side band stabilization} technique, can be found in Refs.~\cite{Drever1983,Black2001}.
	In this work we propose yet another approach to frequency stabilization that employs the effect of gravity on a Fabry-Pérot cavity. In particular, we will use the fact that the cavity output signal depends on the height $h$ in the gravitational field, as discussed in the previous section.
	In the simplest approach this effect can be analyzed by a detector consisting of two, upper and lower, segments, as shown in Fig.~\ref{fig:quad_detector_geometry}. In an experiment, such a detector can be realized on the basis of a common quadrant detector \cite{Diorico2022}.
	Based on Eq.~(\ref{eqn:output_power}) we find that the different vertical positions of the segments' centers translates into their different output signals $P^\pm(\nu)=P_{\mathrm{out}}(\nu,\pm a/2)$.
	In order to quantify this difference we introduce the parameter
	\begin{eqnarray}
		\chi(\nu)&=&\frac{P^{+}(\nu)-P^{-}(\nu)}{P^{+}(\nu)+P^{-}(\nu)}\nonumber\\ &=&
		\frac{ga}{c^2}\,\frac{\pi\nu L}{c} \, \frac{F \sin\left(\frac{4\pi\nu}{c}\,L\right)}{1+F \sin^2\left(\frac{2\pi\nu}{c}\,L\right)}+\order{}(\epsilon^2)\,, \label{eqn:detector_signal}
	\end{eqnarray}
	where we made use of Eq.~(\ref{eqn:output_power}) to obtain the explicit frequency dependence.
	As seen from Eq.~(\ref{eqn:detector_signal}) and Fig.~\ref{fig:quad_detector_signal}, this $\chi$-parameter is zero at resonance frequency $\nu_0$ but changes rapidly for frequencies $\nu=\nu_0+\delta \nu$ close to resonance. 
	In the case of small frequency offsets $\delta \nu$, which is of main interest for frequency stabilization, the $\chi$-parameter can be linearized as
	\begin{equation}
		\chi(\nu_0+\delta\nu)=4\pi^2\,\frac{ga}{c^2}\frac{FL^2\nu_0}{c^2}\,\delta \nu\,\,+\mathcal{O}(\delta \nu^2)\,, \label{eqn:error_function_eq}
	\end{equation}
	which allows us to quantify the capacity of the proposed method for frequency stabilization. In order to illustrate this, let us calculate the linearized $\chi$-parameter for a typical laser-cavity setup.
	We consider a $L=21\,\mathrm{cm}$ cavity with mirrors of reflectivity $R=1-10^{-5}$ coupled to a Nd:YAG-laser, which is widely used for instance in gravitational wave detectors. The wavelength of light produced by this laser is $\lambda= 1064\,\mathrm{nm}$, corresponding to a frequency of $\nu_0=c/\lambda=2.8\times 10^{14}\,\mathrm{Hz}$. Moreover, we assume that the cavity output signal is recorded by a segmented detector as shown in Fig.~\ref{fig:quad_detector_geometry} with $a=1.5 \,\mathrm{cm}$. For such a system, the linearized $\chi$-parameter takes the value
	\begin{equation}
		\chi(\nu_0+\delta\nu)\approx \frac{\delta\nu}{3\,\mathrm{Hz}}\times 10^{-9}\,. \label{eqn:error_function}
	\end{equation}
	To deduce the method's capability for frequency stabilization from this numerical example, one has to know how precise the $\chi$-parameter can be measured. 
	Under ideal experimental conditions, such as optimized measurement frequencies, seismic decoupling and the usage of optical mirror coatings  \cite{Robinson2019,Dooley2015,Notcutt2006}, the sensitivity of the proposed experiment is only limited by the detector photon shot noise 
	\begin{equation}
		\Delta\chi = 0.86\times 10^{-9}\,\, \left(\frac{P_{\mathrm{det}}}{\mathrm{W}}\right)^{-1/2}\,, \label{eqn:photon_noise}
	\end{equation}
	where we considered a common measurement band width of $1\,\mathrm{Hz}$ \cite{Edelstein1978}. Moreover, $P_{\mathrm{det}}=P^{+}(\nu)+P^{-}(\nu)$ is 
	the total detector input power.
	Assuming typical laser powers of a few watts, Eq.~(\ref{eqn:photon_noise}) implies a limit of $\Delta \chi\approx 10^{-9}$.
	This, together with Eq.~(\ref{eqn:error_function}),  implies for the given example that frequency variations up to $3\,\mathrm{Hz}$ can resolved.
	Comparing this result to the used laser frequency of about $3\times 10^{14}\,\mathrm{Hz}$, a relative frequency stabilization in the range of $\delta \nu/\nu\approx 10^{-14}$ can be achieved.
We notice that this is still below the value of $4\times 10^{-17}$ provided by nowadays best frequency stabilization procedures \cite{Kessler2012,Matei2017}.
	However, the achievable frequency stability can be tremendously improved, when in place of the assumed common optical components an optimized experimental setup is used.  
	In particular, since the linearized $\chi$-parameter (\ref{eqn:error_function_eq}) linearly depends on $F=4R/(1-R)^2$, the signal can be enhanced by using higher mirror reflectivities.
	Taking into account the ongoing research in advanced mirror technologies, such as crystalline coatings \cite{Cole2016,Cole2013,Book:Gregory2012} or etalons \cite{Dickmann2018,Somiya2011,Gurkovsky2011,Dickmann2023}, one can consider hypothetical future cavities with $R\sim 1-3\times 10^{-7}$, corresponding to $\mathcal{F}=\pi\sqrt{F}/2=1\times 10^{-7}$.
	In this parameter regime a stabilization procedure based on the CILD-effect can be used to stabilize a laser frequency to the level of $10^{-17}$, as displayed in Table~\ref{tab:detector_signal}.
%
% end.

% new paragraph
The frequency stabilization method presented in this work has some advantages comparing to the previous approaches.
	In particular, the observed $\chi$-parameter does not depend on the laser intensity and, hence, is not sensitive to intensity fluctuations.
	This insensitivity is a common feature of our approach and the side band stabilization technique which constitutes a great advantage over the side-of-fringe stabilization method.
	Furthermore, thanks to the asymmetry of the detector signal (\ref{eqn:detector_signal}), the proposed method allows to distinguish, whether the laser frequency has to be increased or decreased in order to bring it to the cavity resonance frequency $\nu_0$. In turn, 
	this is an advantage over the side-of-fringe stabilization technique, which is based on the analysis of a symmetric signal output. 
	Finally, the presented approach allows to stabilize the laser frequency to $\nu_0$ directly and without the need for frequency modulation, while the other two procedures evaluate either side-band signals or the changes in intensity far off the resonance frequency.
	Together with the relatively simple experimental setup, the advantages mentioned above can make a laser frequency stabilization method based on the CILD-effect suitable for practical applications, especially for large laser powers as used, e.g., in gravitational wave detection.
%
% end.

\section{Summary and outlook} \label{sec:summary}
In this paper we presented a theoretical analysis for the propagation of light within a homogeneous gravitational field.
	The analysis is based on the covariant formulation of Maxwell equations for free electromagnetic fields in a gravitationally distorted (Rindler) spacetime.
	While the developed theory can be applied for various scenarios, here we focus on the propagation of plane waves in a Fabry-Pérot cavity, located in a laboratory on Earth. We found that the output intensity of the cavity resonator is affected by the gravitational field. In particular, it is shown that, due to the cavity internal light deflection effect (CILD-effect) the cavity output intensity not only depends on the frequency of the used laser light but also on the vertical position in the gravitational field. 
 	Based on this effect, we proposed a detector scheme that can be used to stabilize a laser to the cavity resonance frequency.
	More specifically, a vertically segmented detector is proposed to be applied to generate a steep asymmetric output signal close to the cavity resonance.
	Using nowadays state-of-the-art optical components, such as highly reflective mirrors and mono-crystalline silicon, sapphire, or diamond cavities, this method could be competitive with recent approached to frequency stabilization methods in the range of $\delta \nu/\nu = 10^{-15}\dots 10^{-17}$.
	Thus, the CILD-effect would not only enable the verification of gravitational light defection at the laboratory scale but also opens a new alternative way to frequency stabilization which could be applied in modern high precision experiments.

\acknowledgements{The authors would like to thank Stefanie Kroker for helpful discussions. Funded by the Deutsche Forschungsgemeinschaft (DFG, German Research Foundation) under Germany’s Excellence Strategy—EXC 2123 QuantumFrontiers—390837967.}

\begin{table}[t]	
	\caption{\reworked{
			Properties of $\chi(\nu_0+\delta\nu)$ for a segmented detector with $a=1.5 \,\mathrm{cm}$ used to analyze the output intensity of $L=21\,\mathrm{cm}$ cavity with hypothetically high mirror reflectivities of $R=10^{-5}\dots 10^{-7}$ coupled to a laser with $\nu_0=2.8\times 10^{14}\,\mathrm{Hz}$ ($\lambda= c/\nu_0= 1064\,\mathrm{nm}$). The maximum $\chi_{\mathrm{max}}\approx\frac{16}{3}\frac{ga}{c^2} \frac{L\nu_0}{c}\frac{1}{1-R}$ could be measured at a frequency $\nu_0\pm\delta\nu (\chi_{\mathrm{max}})$ off the cavity resonance, where the offset can be estimated $\delta\nu (\chi_{\mathrm{max}})\approx \frac{1}{4\pi}\frac{c}{L}(1-R)$. The inverse steepness of the $\chi$-parameter at resonance $\delta\nu_{\mathrm{min}}=\left.\delta\nu/\chi(\nu_0+\delta\nu)\right|_{\delta \nu=0}\approx \frac{\delta\nu (\chi_{\mathrm{max}})}{2\chi_{\mathrm{max}}}\Delta \chi$ gives the limit for resolvable frequency variations. Calculating the ratio of this value and the resonance frequency $\nu_0$ we obtain an estimate for the relative sensitivity $\delta \nu/\nu\approx \delta\nu_{\mathrm{min}}/\nu_0$. We remark that these values linearly depend on the quadrant detector sensitivity $\Delta\chi$, that is in the range of ppb ($10^{-9}$).\\}}
	\begin{tabular}{p{4em}|p{5em}p{5em}p{5em}p{5em}p{0em}}	\label{tab:detector_signal} \label{page:table}
		%& $\left.\right.$  \centering max. amplitude &  \centering frequency of max. amplitude	& $\left.\right.$ \vfill \centering sensitivity & $\left.\right.$ \vfill \centering \mbox{rel. sensitivity} &	\\[0.1em]
		\centering$1-R$			 & \centering $\chi_{\mathrm{max}}$			&  \centering $\delta\nu (\chi_{\mathrm{max}})$	& \centering $\delta\nu_{\mathrm{min}}$ & \centering $\delta \nu/\nu$ & \\
		& \centering[ppb]  &  \centering[$\mathrm{Hz}$] &	\centering[$\mathrm{Hz}/$ppb] &  & \\[0.5em]\hline \hline
		& 					&											& 	 							& &	\\[-0.6em]
		$1\times 10^{-5}$	 & \centering$172$ 				& \centering $1137$							& \centering$\phantom{0.}3$					& \centering $1\times 10^{-14}$ &	\\
		$3\times 10^{-6}$ 	 & \centering$573$				& \centering$\phantom{0}238$					& \centering$0.2$							& \centering $7\times 10^{-16}$ &  \\
		$1\times 10^{-6}$	 & \centering$1.7\times 10^{3}$	& \centering$\phantom{0}114$				& \centering$3\times 10^{-2}$				& \centering $1\times 10^{-16}$ &	\\
		$3\times 10^{-7}$	 &	\centering$5.7\times 10^{4}$	& \centering$\phantom{00}24$				& \centering$2\times 10^{-3}$				& \centering $7\times 10^{-18}$ &	\\
		$1\times 10^{-7}$	 & \centering$1.7\times 10^{4}$	& \centering$\phantom{00}11$				& \centering$3\times 10^{-4}$				& \centering $1\times 10^{-18}$ & \\[0.2em] \hline
	\end{tabular}
\end{table}	
\begin{appendix}
	
	\section{Power and intensity of light in Rindler spacetime} \label{sec:appendix_A}
	
	\subsection{The general case} \label{sec:light_observables}

	\reworked{To study how light behaves in the homogeneous gravitational field, we need to investigate observables such as energy, power, and intensity. 
		These quantities can be obtained elegantly by using the covariant conservation law 
		\begin{equation}
			\nabla_\mu T^{\mu 0}=0 \label{eqn:conserved_energy}
		\end{equation}
		for the $0$-components of the electromagnetic energy momentum tensor 
		\begin{equation}
			T^{\mu\nu}=\frac{1}{\mu_0}\left(F^{\mu\rho} {F^{\nu}}_\rho-\frac{1}{4}g^{\mu\nu}F^{\rho\sigma}F_{\rho\sigma}\right)\,,
			\label{eqn:em_em_tensor}
		\end{equation}
		cf., e.g., \cite{Book:Carrol2004}.
		By using the definition of the Christoffel symbols in terms of the derivatives of the spacetime metric (\ref{eqn:Metric}), we can rewrite Eq.~(\ref{eqn:conserved_energy}) in the form
		\begin{equation}
			\p_0T^{00}+\frac{1}{\sqrt{-g}}\p_i \left(\sqrt{-g}T^{0i}\right)=0\,,
		\end{equation}
		where $\sqrt{-g}=\left(1+gz/c^2\right)$ is the metric determinant.
		The resulting expression constitutes a continuity equation $\p_t w(\vec{r},t) = -\vec{\nabla}\cdot \vec{S}(\vec{r},t)$ for the energy density and the Pointing vector. Thus, these two quantities are related to the energy momentum tensor by 
		$w(\vec{r},t)=\sqrt{-g}T^{00}$ and $S^{i}(\vec{r},t)/c=\sqrt{-g} T^{i0}$.
	}
	
	\reworked{	% new paragraph
		By integrating the continuity equation over space, we obtain a conservation law in the form $\d E/\d t= P$ relating energy and power in their integral representation:
		\begin{subequations}
			\begin{eqnarray}
				E&=& \int_V\d\vec{r}^3 \sqrt{-g} T^{00}=\int_V\d\vec{r}^3 w(\vec{r},t)\,,\\
				P&=& -\,c\int_V\d\vec{r}^3 \sqrt{-g} \frac{1}{\sqrt{-g}}\p_i\left(\sqrt{-g}T^{0i}\right)\nonumber\\&=&-\int_V\d\vec{r}^3 \vec{\nabla}\cdot\vec{S}(\vec{r},t)=\int_{\p V=\Sigma}\hspace{-2em}
				\vec{S}(\vec{r},t)\cdot\d\vec{\Sigma}\,. \label{eqn:power_definition}
			\end{eqnarray}
		\end{subequations}
		Using Gauss’ integration law in Eq.~(\ref{eqn:power_definition}), we can write the power as an integral over a closed surface $\Sigma$. The resulting statement also holds true for non-closed surfaces, such as detector planes used to analyze the intensity of electromagnetic waves. The integrand $\vec{S}(\vec{r},t)\cdot\d\vec{\Sigma}=I(\vec{r},t)\d\Sigma$ relates the Pointing vector to the intensity $I(\vec{r},t)$ on the surface $\Sigma$.}
	
	\subsection{Application to a vertical detector surface in homogeneous gravity}
	
	\reworked{
		The expression (\ref{eqn:power_definition}) from Appendix \ref{sec:light_observables} gives the definition of power, which is determined by integrating the Pointing vector $S^{i}(\vec{r},t)/c=\sqrt{-g} T^{i0}$ over a specific surface. 
		This surface, representing a detector in experiment, can be chosen as $x=\mathrm{const}$ and proper limits for the $y,z$-integration. 
		For this particular scenario, the infinitesimal surface vector is represented by  $\d\vec{\Sigma}=\vec{e}_x\,\d y\d z$ and the integrand of the power integral
		\begin{eqnarray}
			P&=& \int_{z_1}^{z_2}\int_{y_1(z)}^{y_2(z)} S^1(\vec{r},t) \,\d y \d z\\
			&=& \int_{z_1}^{z_2}\int_{y_1(z)}^{y_2(z)} \sqrt{-g} \,c\,T^{10} \,\d y \d z \label{eqn:output_power_definition}\,,
		\end{eqnarray}
		can be identified as the intensity $I(\vec{r},t)= \sqrt{-g} \,c\,T^{10}$. 
		When, for instance, a $y$-polarized plane wave vector potential $\vec{A}(\vec{r},t)$ from Eq.~(\ref{eqn:Art}) is assumed, the intensity can be further simplified	
		\begin{eqnarray}
			I(\vec{r},t)
			&=&-\frac{1}{2}\varepsilon_0 c^2\left(1+\frac{gz}{c^2}\right)^{-1}\, \mathrm{Re}\bigl[\p_t\vec{A}(\vec{r},t)\cdot  \p_x\vec{A}^{\ast}(\vec{r},t)\bigr]\nonumber\\
			&=&\frac{1}{2}\varepsilon_0 c\,(2\pi\nu_0)^2 \left(1+\frac{gz}{c^2}\right)^{2\alpha-\beta-3}|\vec{A}_0|^2\,. \label{eqn:intensity}
		\end{eqnarray}
		The appearance of the term $(1+gz/c^2)$ in the equation indicates that an height-dependent intensity damping may occur depending on the initial conditions of the experiment. However, this damping only becomes relevant when intensities are measured with a relative precision of $gL/c^2\sim 10^{-18}$ for $L\sim 1\mathrm{cm}$, which exceeds the precision of current intensity measurements techniques by nine orders of magnitude \cite{Edelstein1978}. 	
		We, therefore, can assume a constant intensity profile $I_0=\varepsilon_0 c\,(2\pi\nu_0)^2 |\vec{A}_0|^2/2$ for plane waves along a detector surface at $x=\mathrm{const}$ within the technical capabilities of current experimental equipment.
	}
\end{appendix}

%*******************************************************
% Bibliography
%*******************************************************

\newcommand{\tit}[1]{\emph{#1}}
\newcommand{\owntit}[1]{\textbf{#1}}
\newcommand{\vol}[1]{\textbf{#1}}

\newcommand{\CQG}[1]{Classical and Quantum Gravity \vol{#1}}
\newcommand{\PRL}[1]{Physical Review Letters \vol{#1}}
\newcommand{\PRD}[1]{Physical Review D \vol{#1}}
\newcommand{\PRA}[1]{Physical Review A \vol{#1}}
\newcommand{\PRX}[1]{Physical Review X \vol{#1}}
\newcommand{\PR}[1]{Physical Review \vol{#1}}
\newcommand{\PLA}[1]{Physics Letters A \vol{#1}}
\newcommand{\PS}[1]{Physica Scripta \vol{#1}}
\newcommand{\PhysRep}[1]{Physics Reports \vol{#1}}
\newcommand{\PTEP}[1]{Progress of Theoretical and Experimental Physics \vol{#1}}
\newcommand{\RMP}[1]{Reviews of Modern Physics \vol{#1}}
\newcommand{\NPhoto}[1]{Nature Photonics \vol{#1}}
\newcommand{\NPhys}[1]{Nature Physics \vol{#1}}
\newcommand{\Nature}[1]{Nature \vol{#1}}
\newcommand{\NatCom}[1]{Nature Communications \vol{#1}}
\newcommand{\NJP}[1]{New Journal of Physics \vol{#1}}
\newcommand{\Metro}[1]{Metrologica \vol{#1}}
\newcommand{\Opt}[1]{Optica \vol{#1}}
\newcommand{\OL}[1]{Optics Letters \vol{#1}}
\newcommand{\OC}[1]{Optics Communications \vol{#1}}
\newcommand{\OptExp}[1]{Optics Express \vol{#1}}
\newcommand{\EPL}[1]{Europhysics Letters \vol{#1}}
\newcommand{\EPJST}[1]{European Physical Journal Special Topics \vol{#1}}
\newcommand{\AdP}[1]{Annalen der Physik \vol{#1}}
\newcommand{\JPC}[1]{Journal of Physics Communications \vol{#1}}
\newcommand{\JPCS}[1]{Journal of Physics: Conference Series \vol{#1}}
\newcommand{\JPESI}[1]{Journal of Physics E: Scientific Instruments \vol{#1}}
\newcommand{\PTRSL}[1]{Philosophical Transactions of the Royal Society of London. Series A, \vol{#1}}
\newcommand{\IEEE}[1]{IEEE Transactions on Ultrasonics, Ferroelectrics, and Frequency Control \vol{#1}}
\newcommand{\AJP}[1]{American Journal of Physics \vol{#1}}
\newcommand{\MolPhys}[1]{Molecular Physics \vol{#1}}
\newcommand{\APB}[1]{Applied Physics B \vol{#1}}
\newcommand{\AO}[1]{Applied Optics \vol{#1}}
\newcommand{\Science}[1]{Science \vol{#1}}
\newcommand{\SM}[1]{Sidereal Messenger \vol{#1}}
\newcommand{\ZP}[1]{Zeitschrift für Physik \vol{#1}}
\newcommand{\ProcRS}[1]{Proceedings of the Poyal Society A \vol{#1}}


\begin{thebibliography}{[10]}
\bibitem{remark1}
	\approved{We notice that the intensity (\ref{eqn:output_intensity}) is invariant under vertical
		shifts of the coordinate center $z'=z-h$, when simultaneously the frequency is replaced by the corresponding redshifted frequency $\nu'=\nu\left(1+gh/c^2\right)^{-1}$, such that
		$I_{\mathrm{out}}(\nu',z')=I_{\mathrm{out}}(\nu,z)+\mathcal{O}(\epsilon^2)\,$.
		This implies that expression (\ref{eqn:output_intensity}) does not depend on the initial choice of the coordinate system described by the metric (\ref{eqn:Metric}) and is suitable to describe the intensity profile art arbitrary height $h$ in the homogeneous gravitational field.}
\end{thebibliography}
\end{document}